**Statistical Methods for Accommodating Immortal Time: A Selective Review and Comparison**

Jiping Wang[1], Peter Peduzzi[1], Michael Wininger[2,1], Shuangge Ma[1*]

[1]Department of Biostatistics, Yale School of Public Health, New Haven, CT, USA

[2]Cooperative Studies Program Coordinating Center, VA Connecticut Healthcare System, West Haven, CT, USA

*For correspondence: shuangge.ma@yale.edu

**Abstract**

Epidemiologic studies and clinical trials with a survival outcome are often challenged by immortal time (IMT), a period of follow-up during which the survival outcome cannot occur because of the observed later treatment initiation. It has been well recognized that failing to properly accommodate IMT leads to biased estimation and misleading inference. Accordingly, a series of statistical methods have been developed, from the simplest by including or excluding IMT to various weightings and the more recent sequential methods. Our literature review suggests that the existing developments are often "scattered", and there is a lack of comprehensive review and direct comparison. To fill this knowledge gap and better introduce this important topic especially to biomedical researchers, we provide this review to comprehensively describe the available methods, discuss their advantages and disadvantages, and equally important, directly compare their performance via simulation and the analysis of the Stanford heart transplant data. The key observation is that the time-varying treatment modeling and sequential trial methods tend to provide unbiased estimation, while the other methods may result in substantial bias. We also provide an in-depth discussion on the interconnections with causal inference.

**1. Introduction**

Immortal time (IMT) bias is frequently encountered in epidemiologic studies and clinical trials that involve a survival outcome and comparison of two or more groups (for example, two treatment strategies). In the literature, it has also been termed length-sampling bias,[1,2] survivor treatment bias,[3] time-dependent bias,[4] and survivorship bias.[5] IMT bias is an information bias,[6,7] resulting from a follow-up period during which the survival event of interest cannot occur.[8] This bias results when the data at study entry is insufficient to determine which treatment strategy an individual's data is consistent with (for example, a subject has not initiated treatment when follow-up starts), and is tightly connected with ignoring or improperly dealing with the time-varying nature of treatment status.[3] Below we first describe the IMT challenge in observational studies and clinical trials.

**Observational studies** Generically, we consider an observational study that compares survival between a treatment and a control group. Such a study has three important time points: time of meeting eligibility, time of starting follow-up, and time of treatment initiation. In a typical observational study, the time of starting follow-up is "missing", and the three time points are usually not sufficiently close. A common practice is to use the time of the first identification of eligible disease conditions as time zero, and the first subsequent observed treatment status (before the survival event) for treatment group classification. This was the case in the 1970s when IMT bias was first noted in the analysis of the Stanford Heart Transplant Program data generated from a prospective cohort study to evaluate the effect of cardiac transplantation for patients with end-stage heart diseases.[9] It was recognized that the original control group of this study was not properly chosen, as the principal reason that a candidate did not receive transplantation was that he/she died before a donor was available.[10] The observed shorter survival of the control group gave an artificial boost to the relative survival of the transplanted patients. We refer to published studies for analysis of these data as well as extensive discussions on IMT bias.[10-12] In the seminal paper by Gail,[9] the span from time zero to the time a patient received his/her transplantation was first termed as a "grace period" or "waiting period", and later more commonly referred to as "immortal time". More recent examples include studies of chronic obstructive pulmonary disease,[13-17] asthma,[18,19] AIDS,[20] device implantation for heart failure,[21] surgery for cataracts,[22,23] transplantation for acute lymphoblastic leukemia,[24] hematopoietic malignancies,[25] and pollution exposure.[26-28]

**Clinical trials** In contrast to observational studies, in clinical trials we know accurately the time of meeting eligibility, time of starting follow-up (random treatment assignment), and time of treatment initiation – there is no "missingness". However, if the three time points cannot be combined into one time zero or if participants deviate from the originally assigned treatments after the start of follow-up,[29] IMT may arise in the per-protocol (PP) analysis,[30] which may be more appropriate than the intention-to-treat (ITT) analysis when there is moderate to severe non-adherence.[31-33] A representative study that examined non-adherence and IMT is Peduzzi et al. (1993),[34] which analyzed data from the Veterans Affairs Cooperative Study of Coronary Artery Bypass Surgery.[35] This trial was designed to compare the survival of Veterans assigned to optimal medical therapy versus those assigned to coronary artery bypass surgery. However, nearly half of the participants in the control group did not adhere to their original treatment assignment and had bypass surgery (i.e., crossovers) by the end of 14 years of follow-up. The waiting time for crossovers ranged from 4 to 8 years, and this waiting time was the IMT during which death could not occur. In these types of studies, IMT is a consequence of non-adherence and needs to be properly addressed.

Since the 1970s, it has been well recognized that IMT widely exists in observational studies and clinical trials, and that failing to properly account for it can lead to biased estimation of treatment effects and misleading inferences. Accordingly, a myriad of statistical techniques have been developed. A partial methodological development timeline is shown in Figure 1. Early efforts in the 1970's started with addressing the time-varying feature of treatment status by adopting dynamic group definitions and statistical test procedures. Over the next two decades, there was more attention to modeling (missing) time zero, with approaches such as landmark analysis[36] and Prescription Time Distribution Matching (PTDM).[37] Beginning around 2000, much effort was devoted to integrating emulation analysis with duplication and a nested trial design.[38, 39] Several very recent studies, such as Maringe et al. (2020),[40] Thomas et al. (2020),[41] and Harding and Weiss (2019),[23] suggest that IMT research still remains very active. As discussed below, the existing methods take highly different strategies and have different working characteristics.

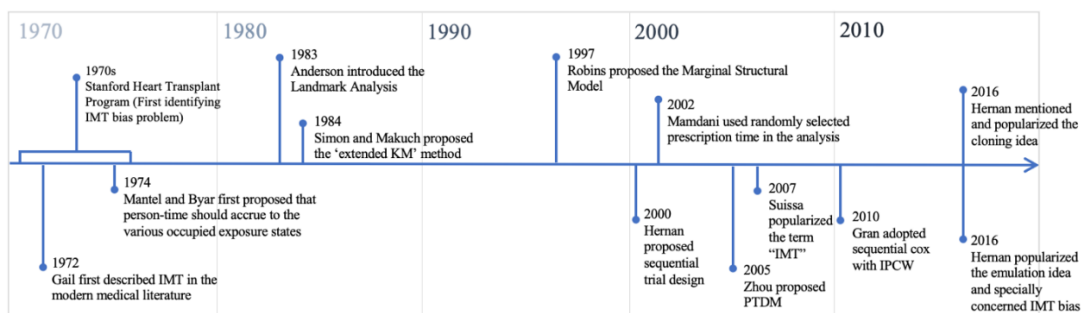

*Figure 1 A partial timeline of methods on immortal time bias*

The objectives of the present study are multi-fold. The first is to provide a selective review of the existing methods for accommodating IMT and with the goal of removing/reducing bias. Our literature review suggests that most of the existing studies have focused on the development of a single approach; however, there is a lack of study that comprehensively reviews multiple approaches, positions one against another, properly classifies them, and discusses their advantages and disadvantages.[39, 40, 42, 43] Because IMT is likely to exist in observational and clinical trial studies, biomedical researchers need to be familiar with the existing methodologies. The second objective is to conduct numerical analysis and directly compare different approaches. Such an effort is missing in the literature. Some of the existing studies have discussed performance of different approaches without providing objective numerical support,[8, 21, 44, 45] while some others have selectively compared a few approaches in limited settings.[46, 47] Our numerical study can complement the existing literature by providing direct, more comprehensive, and more objective comparisons, which can effectively inform users on choosing proper analysis methods. The third objective, which also fills a knowledge gap, is to provide a thorough discussion on "combining" methods for IMT, confounding bias, and causal inference. To also accommodate biomedical researchers who may not have extensive statistical training, we will focus on the intuition and operation of methods, as opposed to mathematical details and theories.

## 2. Review of methods

### 2.1 Settings and notations

Consider a study with a time-to-event response as the designated endpoint and two treatment arms (referred to as treatment and control). The analysis of more complex settings can follow using extensions of the reviewed methods. It is assumed that a subject cannot initiate treatment before entering the study, and that, once treatment is initiated, a treated subject cannot switch back to the control group. Consider a sequence of $K$ time intervals, with the $k$th time interval starting at $t_k$ and ending at $t_{k+1}$. $t_1 = 0$ denotes the time of study entry (first meeting eligibility). For subject $i$ ($i = 1, ..., n$), $L_{k,i}$ denotes the covariate vector observed at time $t_k$. Subject $i$ may initiate treatment in any time interval, and the true treatment indicator at time interval $k$ is denoted as $A_{k,i} = 1$ (treatment) or $0$ (control). Denote $T_{A,i}$ as the exact treatment initiation time. Subject $i$ is followed until the time of event $T_{F,i}$, loss to follow-up $T_{C,i}$, or $t_{K+1}$, whichever occurs first. The censoring indicator $C_i = 1$ if $\min(T_{F,i}, T_{C,i}, t_{K+1}) = T_{F,i}$, and $C_i = 0$ otherwise.

We also define method-specific notations. Denote $T'_i$ as time zero for analysis. If time zero has the same definition for all subjects in the treatment (control) group, it is denoted as $T'_{trt}$ ($T'_{ctrl}$). Some methods do not model time-varying treatment status but rather assign a time-independent treatment indicator for each observed (or duplicated) record. In this case, the assigned treatment is denoted as $A'_i$.

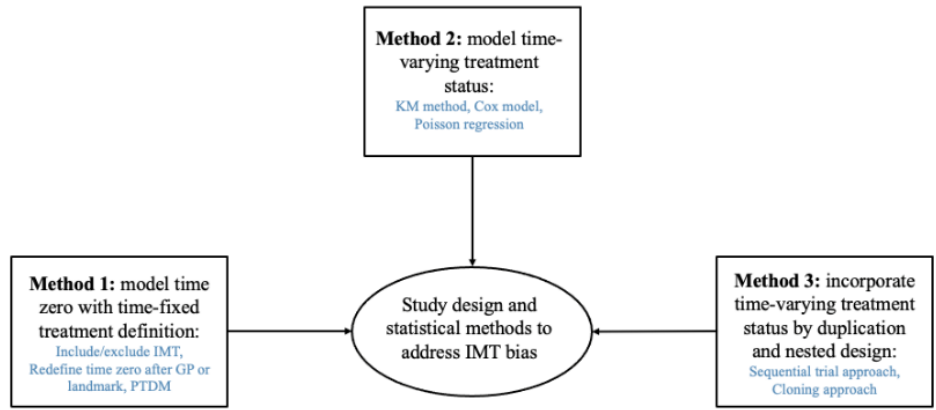

*Figure 2 Classification of methods*

**2.2 Classification of methods**

To address IMT, researchers have examined time zero, treatment assignment, and overall study design. We classify the existing methods into three families (Figure 2) and note that the classification is not absolute, and there can be methods "in the middle". (i) The first family of methods uses an alternative time zero, such as the time of treatment initiation as opposed to study entry. A subject's treatment classification is time-independent, usually based on an ever- or never-treated definition. (ii) The second family is characterized by modeling time-varying treatment status, based on which various model fitting can be conducted. (iii) The third family, as represented by the cloning and sequential trial methods, involves creating one or multiple copies of one subject and assigning copies to the two arms.

To more intuitively describe each method, we consider a representative exemplar, as depicted in Figure 3. Figure 3(a) describes the ideal situation where the time of study entry and time of treatment initiation coincide at an unambiguous time zero. The four subjects in the treatment arm (green shaded region) and those in the control arm (red shaded region) have the same survival times, and thus there is no treatment effect. Figure 3(b) describes the more realistic situation in which treatment is initiated (represented by green dots) after study entry. It is noted that subject 3 has developed the event of interest (represented by black dots) before treatment can initiate. From visual inspection of Figure 3(b), it is clear: the treatment and control arms no longer have the same survival time distribution if either subject 3 is classified as in the control arm or the study entry is set as the time of treatment initiation.

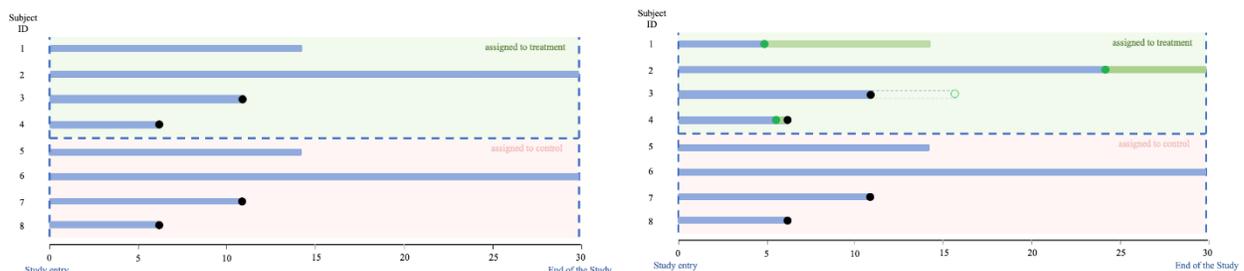

*Figure 3 Demonstration of immortal time. (a) left: ideal data, as might be obtained in a clinical trial, where treatment initiation coincides with time of enrollment; (b) right: naturalistic data, as might be obtained through observational study, where treatment initiation does not necessarily coincide with time of enrollment (green dots at time beyond $t_1$).*

## 2.3 Methods that model time zero

This family of methods mainly includes those misclassifying/including IMT, those excluding IMT, PTDM, as well as landmark analysis and redefining time zero after grace period.

### 2.3.1 Including IMT

This method follows the original problematic study design and may incur the most severe bias. For all subjects enrolled in the study, time zero is defined as the study entry $T' = t_1 = 0$. Those who have ever initiated treatment are classified as in the treatment group ($A'_i = 1$), and those who have never initiated treatment during follow-up are classified as in the control group ($A'_i = 0$). Extending the exemplar shown in Figure 3, the data for analysis is presented in Figure 4 (a). Subjects 1, 2, and 4 form the treatment group, time zero is defined as the beginning of the 0$^{th}$ month, and all subjects are followed until censoring or event. For example, subject 1 is followed from the 0$^{th}$ month to the end of the 14$^{th}$ month.

A real-world example is offered in Sin et al. (2002)[48], which evaluated the effects of beta-blockers on preventing morbidity and mortality for patients with heart failure. All subjects were followed from the date of their index hospitalization until the outcome or censoring, and all the dosage and group classification were determined after the start of follow-up. Bias in estimation was caused by using information obtained after baseline to determine treatment assignment at baseline, resulting in misclassification of the proportion of person-time (from $t_1$ to $T_{A,i}$) as exposed when the treatment had not yet been initiated. The artificial increase in person-time of the treatment group can explain the systematically underestimated exposure hazards or boosted treatment effect.[46] The researchers further performed a sensitivity analysis in which they excluded patients who died within 30 days of the index hospitalization. Gleiss et al. (2018)[49] and Zhou et al. (2005)[37] pointed out that a longer IMT between study entry and treatment initiation can lead to more bias, and the magnitude of bias can also depend on the mortality risk before and after treatment initiation.

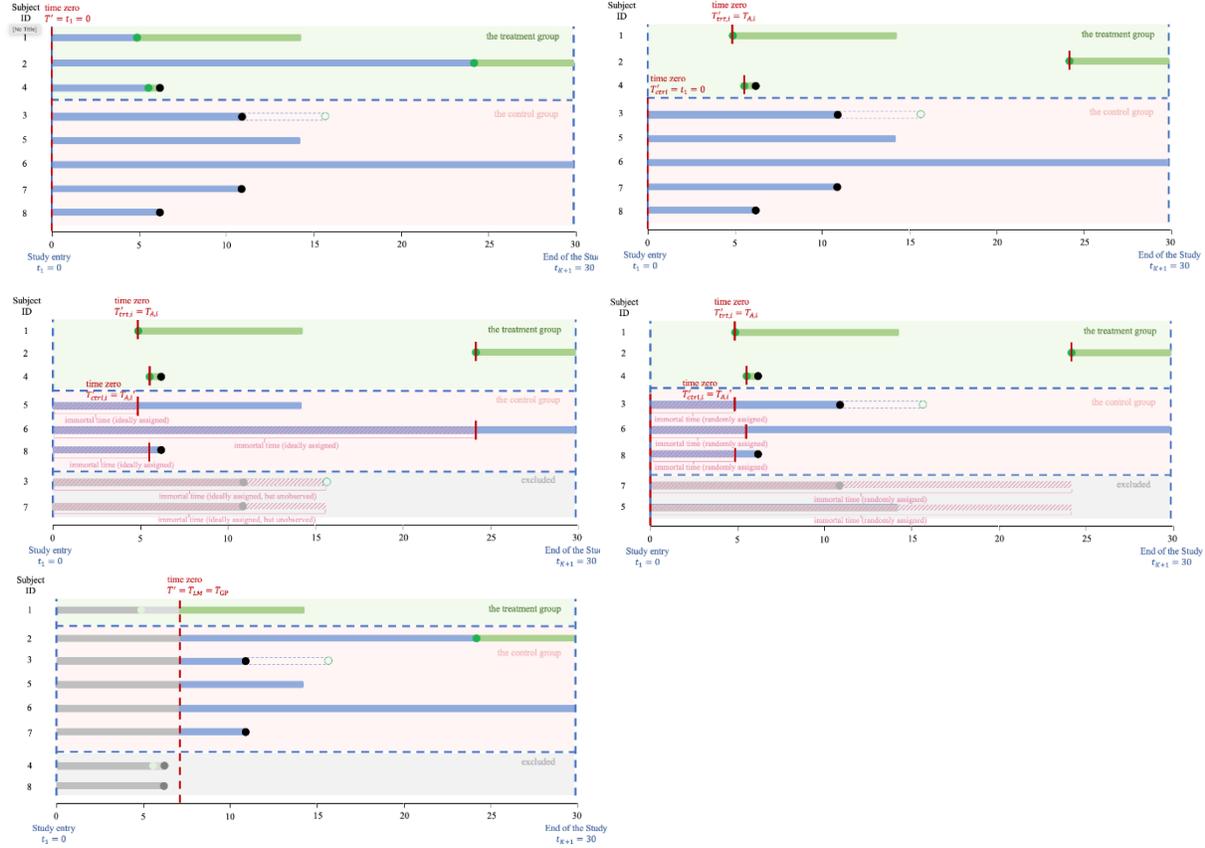

*Figure 4 Various approaches to accommodating immortal time bias. (a) left-upper: including IMT; (b) right-upper: excluding IMT; (c.1) left-middle: PTMD, the ideal case where matching is exactly correct and unobserved IMT information is available; (c.2) right-middle: the real-world case; (d) left-lower: landmark analysis and redefining time zero after grace period.*

### 2.3.2 Excluding IMT

The key difference between this approach and the one in Section 2.3.1 is that time zero for the treatment group is no longer study entry. Specifically, for those who have ever initiated treatment ($A'_i = 1$), IMT is excluded from analysis, that is, time zero is redefined as the time point they start treatment $T'_{trt,i} = T_{A,i}$. For the control group ($A'_i = 0$), time zero and treatment assignment are the same as in Section 2.3.1 ($T'_{ctrl} = t_1 = 0$). For the exemplar (Figure 4 (b)), subjects 1, 2, and 4 are classified as in the treatment group. Time zero for the treated is the time of treatment initiation, and so the follow-up time of subjects 1, 2, and 4 after treatment initiation contributes to the treatment group before they reach censoring or event. In the literature, it has been recognized that this method may correctly estimate the hazard for the treatment group but not the control group, and that bias is larger when more IMT is excluded.[21, 50]

As an example, this method was adopted for evaluating the treatment effect of revascularization.[51] For patients undergoing revascularization, follow-up started at the time of initial intervention, while for those not receiving revascularization, it started at the time of initial catheterization (study entry).

### 2.3.3 PTDM

Researchers have targeted the missing waiting time/IMT components for subjects in the control group and resetting time zero as study entry plus the assigned IMT. Treatment classification is based on the ever-treated and never-treated status across entire follow-up. Time zero for the treatment group is the time point of treatment initiation $T'_{trt,i} = T_{A,i}$, and for the control group, it is the randomly imputed IMT. For example, Mamdani et al. (2002)[52] randomly selected a possible day in the grace period (a special time window after which continuation of treatment and no new treatment initiation are assumed) as the IMT for the control group. However, this approach may not lead to an effective estimation of the survival difference between the two arms, as the uniform distribution of IMT for the control group may severely deviate from the distribution observed in the treatment group. To tackle this problem, Zhou et al. (2005)[37] proposed PTDM, which randomly imputes the same missing piece but from the real distribution observed on the ever-treated subjects. Specifically, the observed IMTs for the ever-treated subjects are first computed. The matched IMTs for the control group are selected randomly from these values with replacement, and as such, IMT has the same distribution for the treatment and control groups. For example, time zero for subject $i$ in the control group is $T'_{ctrl,i} = T_{A,i'}$, where $T_{A,i'}$ is the observed IMT for subject $i'$, who is matched with subject $i$. If subject $i$ experiences an event before the end of the assigned IMT, he/she is excluded from further analysis. In reference to the exemplar, the scheme of this analysis is shown in Figure 4 (c.1) and (c.2). In an ideal world, we would know that subjects 1 and 5, 2 and 6, 3 and 7, as well as 4 and 8 are matched, and be aware of unobservable information like IMT for those never-treated subjects. As shown in Figure 4 (c.1), we would use the matched IMT (either observed or unobserved) to redefine time zero for subjects in the control group. For subject pairs 1 and 5, 2 and 6, as well as 4 and 8, the IMT values would be matched. As treatment initiation would happen after the developed outcome, subjects 3 and 7 would be excluded from analysis – if the ideally assigned but unobservable IMT was known. In contrast, in the real world (Figure 4 (c.2)), we do not know the matched pairs perfectly and do not have unobserved information. As such, the never-treated subjects may be mismatched. For example, subject 6 may be matched with subject 4, and subjects 5 and 7 are excluded from analysis.

Keyhan et al. (2007)[53] reported a real-world application of PTDM for angiotensin-converting enzyme (ACE) inhibitors usage for congestive heart failure under a retrospective cohort design with over 27,000 participants. For the ever-treated group, participants were followed after the first prescription. And for the never-treated group, participants were followed after the randomly imputed time zero (which lies between hospital discharge and the first prescription). Only a small proportion (482/8617) of the never-treated who had event before the assigned time zero were excluded from analysis. An alternative of PTDM is matching not only on prescription time but also baseline covariates, given the matched control is alive at the time when the treated subject initiates treatment. For example, Pittet et al. (1994)[54] examined the effect of nosocomial bloodstream infection using a matched cohort of critically illed patients. In this study, each patient with a bloodstream infection (exposed) was matched to a control patient based on patients' characteristics and IMT before the onset of bloodstream infection, and the outcome was the days remaining in ICU. Those who developed outcome were excluded for matching. Therefore, both the exposed and controls had the same IMT distribution, and so IMT bias could be avoided. A major limitation of this method is the distortion of the target population and a loss of statistical efficiency and precision.[37, 55, 56] Specifically, it excludes never-treated subjects with certain features (developing outcomes early). The altered target population is prone to those more likely to initiate treatment and those in the control group who have a longer event-free follow-up.

*2.3.4 Landmark analysis and redefining time zero after a grace period*

Instead of using the time of eligibility or treatment initiation as imputed time zero, some observational studies circumvent the IMT problem by redefining a clinically relevant time zero. [49, 57] Specifically, a fixed time point ($T_{LM}$) after the initiation of treatment is selected as the landmark, and time zero is defined as this specified landmark ($T' = T_{LM}$). Subjects who are under treatment at $T_{LM}$ are classified as treated, and the others are classified as controls. As such, group assignment depends on not only the ever- and never-treated status but also the time of treatment initiation $T_{A,i}$ and choice of landmark $T_{LM}$. Subjects who experience event or are censored before the landmark are excluded from analysis. The remaining subjects are followed from $T_{LM}$, subsequent switches in treatment are ignored, and survival difference is quantified based on treatment information at the landmark.

A grace period (ending at $T_{GP}$) is a special case of the time window from study entry $t_1 = 0$ to landmark $T_{LM}$, after which continuation of treatment and no new treatment initiation are assumed.[15, 58] Similarly, group status ($A'_i$) for analysis is determined based on time-varying treatment status during the grace period: $A'_i = 1$ if treatment is initiated during the grace period and 0 otherwise.

The exemplar under landmark analysis and grace period analysis is shown in Figure 4 (d). If we for example set the landmark time as the 6th month, only subject 1 initiates treatment before the reset time zero and thus is classified as treated. All the other subjects are classified as control, regardless of their later treatment initiation. Subjects 4 and 8 experience event before the landmark time and are excluded from further analysis.

An example is a cohort study on the effect of statins on AMI recurrence or all-cause mortality for elderly patients who were discharged alive with a diagnosis of acute myocardial infarction.[37] Statin users and non-users were identified based on treatment status within 90 days post-discharge and followed from the end of the exposure time window (i.e., 90 days post-discharge) until the occurrence of outcome. Similar to PTDM, landmark analysis and grace period analysis may also cause distortion of the target population and a loss of statistical efficiency and precision.[37, 55, 56] For example, Herzog et al. (2005)[58] conducted a retrospective cohort study on the survival impact of Implantable Cardioverter Defibrillators (ICDs), which identified dialysis patients with or without ICD implantation within 30 days of hospital admission after cardiac arrest and followed patients from day 30 after admission. In this study, only 26% (7,853/30,230) of the patients were alive at the redefined time zero, and a significant proportion of potentially eligible patients were excluded from analysis. The sample consisting of those alive at time zero no longer represents the initial eligible cohort at hospital admission (the population of interest). Furthermore, the reduced sample size leads to reduced power. Another limitation of grace period analysis is that it is not applicable to some scenarios. For example, when the switch of treatment happens frequently, change of treatment status should be monitored through the entire follow-up period, and grace period can be too long and accounts for a "non-negligible" proportion of follow-up.

Grace period-based analysis, loosely speaking, is a special case of landmark analysis. However, there are notable differences. In landmark analysis, the assumptions of continuation of treatment and no new treatment initiation after a certain period is not stressed. Except for using the end of grace period as a landmark, the treatment effect derived from other landmark choices focuses on the average survival of subjects conditional on being event-free and on the treatment status assessed at landmark. This interpretation of the derived treatment effect estimate may restrict applicability of this method.[24] In addition, since choosing a landmark can be difficult and subjective, sensitivity analysis with a sequence of landmarks is often needed. Researchers have also built supermodels based on multiple landmarks.[49, 59] For example, Haller et al. (2017)[59] evaluated the effect of steroid withdrawal at various time points after kidney transplantation on patient and graft survival. A series of landmark points were considered. At each landmark point, the study participants were classified as steroid-withdrawal or steroid-maintenance depending on steroid treatment status within the preceding time interval. The researchers then resorted to propensity score matching and calculated the cause-specific cumulative incidence functions for the competing events at specific landmark times. To get the final estimate, they estimated a landmark-stratified Cox supermodel using all matched data from all landmarks. We note that this supermodel is somewhat similar to the sequential trial approach to be discussed in Section 2.5.1.

**2.4 Methods that model time-varying treatment status**

This family of methods is characterized by time-varying treatment status, which can be coupled with Kaplan-Meier estimation,[24, 42, 60] Cox model,[12, 46, 61, 62] Poisson model,[62] and others. Modeling treatment status as real-time and time-varying has been considered as the gold standard for handling IMT as it provides almost unbiased estimation if done properly.[21, 46] Specifically, a time-dependent variable for treatment initiation $A_{k,i}$ ($k = 1, ..., K$) is used to define treatment and control status in time interval $k$. It is assumed that once treatment is initiated, a subject will adhere to the assigned treatment. Time zero is no longer applicable as treatment assignment is not done based on a single time point. Follow-up starts at study entry (baseline) $t_1 = 0$ and ends at $T_{F,i} = \min(T_{F,i}, T_{C,i}, t_{K+1})$. As demonstrated using the exemplar in Figure 5, this method accurately represents treatment status and classifies event-free person-time of ever-treated subjects before their treatment initiation as follow-up time of the control group.

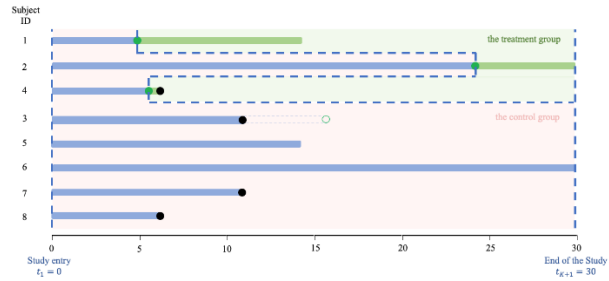

*Figure 5 Exemplar demonstrating methods that model time-varying treatment status*

Many studies have adopted time-varying modeling to derive treatment effect.[10, 12, 63, 64] For example, a prospective cohort study used both time-fixed and time-varying Cox models to analyze the association between delirium in the ICU and ICU length-of-stay.[63] The time-fixed analysis yielded an apparently statistically significant exposure effect, while the time-varying analysis did not. Simulation was conducted to quantify bias and showed that time-varying modeling tended to minimize bias in estimation while time-fixed modeling results were heavily biased.

This method also has limitations. First, with some analysis approaches such as propensity score adjustment and emulation, treatment status must be defined at time zero, and thus the time-varying definition of treatment status cannot be applied. Second, this method is not conveniently implementable using "traditional" software and packages. Also, if we adopt parametric or semi-parametric models, data needs to be prepared in a special longitudinal format under which a subject's follow-up is broken into multiple segments to record every treatment initiation,[65] which leads to extra storage and computation burden. Third, the group-based interpretation is tricky under the time-varying treatment status definition.[49, 66, 67] For example, patient characteristics can change over time and at the time of treatment change. Also, data are often not available for everyone at the times of treatment change. Last, with the dynamic treatment group definition, Kaplan–Meier estimation and related techniques may be less appropriate (compared to parametric and semi-parametric modeling).[49]

## 2.5 Methods that involve duplication and nested design

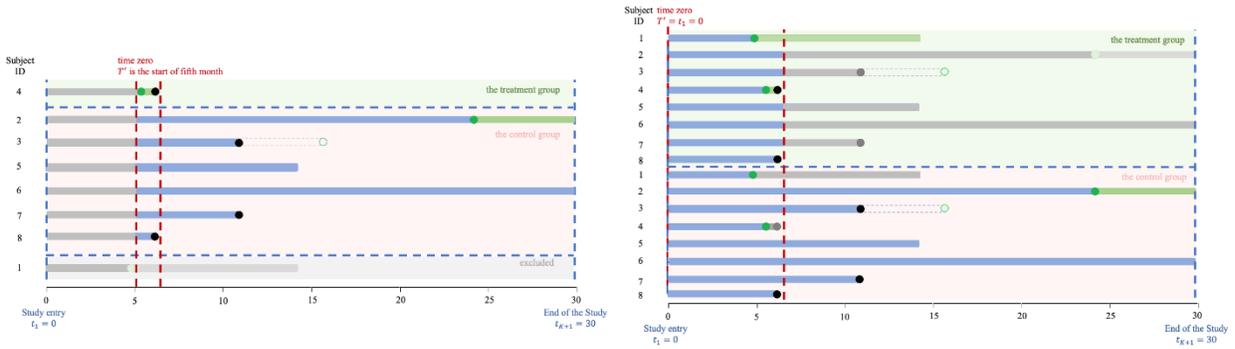

*Figure 6 Exemplar demonstrating methods that involve duplication and nested design. (a) Left: sequential trial approach; (b) right: cloning approach.*

In this family, the sequential trial approach models differentiated IMT by a sequence of time zeros and thus can account for time-varying treatment status. With the cloning approach, two copies of each subject are separately assigned to two treatment arms, and follow-up is modeled based on true adherence, making it possible to accommodate time-varying treatment assignment.

### *2.5.1 Sequential trial approach*

One way to account for time-varying treatment status is to incorporate all potential time zeros and pool all estimations together. Both subject-specific IMT (waiting time) and sojourn time (from treatment initiation to endpoint)[24] are modeled by sequential and nested trial designs. Specifically, the sequential trial approach mimics a series of randomized controlled clinical trials or cohort studies, and each trial (study) starts in one time interval with a potential of treatment initiation within that interval.[43] For the $k$th trial (study), we only include those who are eligible at baseline $t_1 = 0$, survive at least until $t_k$ ($T_{F,i} > t_k$), and have not initiated treatment until $t_k$ ($T_{A,i} \geq t_k$ or $T_{A,i} = $ NA). Time zero is set as $t_k$ (or another point within the time interval $[t_k, t_{k+1})$). Treatment assignment is based on whether a subject initiates treatment in $[t_k, t_{k+1})$. In particular, those who initiate treatment within $[t_k, t_{k+1})$ are classified as in the treatment group for trial $k$, and those who have not initiated treatment by $t_{k+1}$ ($T_{A,i} \geq t_{k+1}$ or $T_{A,i} = $ NA) are classified as in the control group (regardless of whether they initiate treatment after $t_{k+1}$). This process is repeated for each trial, and all sequential trials are pooled. If the stratified Cox model is adopted (stratified by each trial),[68, 69] the assumptions are that subjects enrolled in different trials have different baseline hazards, all subjects enrolled in the same trial have the same baseline hazards, and all trials have the same treatment effect.

If the abovementioned same treatment effect assumption is not satisfied, we may interpret estimation as an "average" over all trials.[43] Similarly, we can use pooled logistic regression to derive odds ratio which mimics the hazard ratio.[70, 71] It is noted that since one subject can be included in multiple trials, a robust sandwich estimator or bootstrap is needed for variance calculation to account for the correlation.[46, 72, 73] Further, controls in a trial may later initiate treatment after the initial enrollment window. If we neglect later deviation from the protocol, the derived treatment effect is more like an ITT estimate, and the observed treatment statuses at the baseline of each trial are our intention-to-assigned. If we are more interested in PP treatment effect, we need to further censor those subjects when non-adherence happens and weighting methods are potentially needed to address selection bias caused by artificial censoring.[33, 74] For relevant discussions, we refer to Gran et al. (2010),[43] Karim et al. (2016),[46] and Danaei et al. (2018).[71]

For the exemplar, the ITT analog using the sequential trial approach is shown in Figure 6 (a). We take the 5$^{th}$-month trial as an example. Those who develop outcome, initiate treatment, or are censored before the beginning of the 5$^{th}$ month are excluded from this trial. For example, subject 1 is excluded. Subject 4 is the only one that initiates treatment during the one-month enrollment window (second dotted line) and is classified as treated. The other subjects form the control group. For ITT analysis, although subject 2 initiates treatment later after the enrollment window and deviates from the originally assigned treatment, we ignore the deviation and assume adherence until outcome or censoring is observed. For PP analysis, we censor subject 2 at the time of protocol deviation and apply weighting to address informative censoring.

An application example is on estimating statin's effect on primary prevention of coronary heart disease,[74] which resorted to this approach to increase sample size. More recently, Gupta et al. (2021)[75] studied the survival of COVID -19 patients with early Tocilizumab treatment. The sequential trial technique served as a sensitivity analysis to evaluate the impact of IMT on the estimated treatment effect.

*2.5.2 Cloning approach*

In a sense similar to the sequential trial approach, the cloning approach also deals with IMT by duplicating observations.[39] The differences are that the cloning approach creates exactly two copies of each observation (with one copy for each treatment group) and sets the same time zero for all copies and observations. This approach has three main steps: (i) Clone observations, and assign them to the treatment and control groups. Subject $i$ is cloned into two copies $j_{trt,i}$ and $j_{ctrl,i}$, which have the same baseline covariates. Copy $j_{trt,i}(j_{ctrl,i})$ is assigned to the treatment (control) group, regardless of later treatment initiation. Both copies start being followed at $T'_i = t_1 = 0$. (ii) Censor clones when they deviate from their assigned treatments. Here we need to define the grace period (usually) based on clinical practice.[40] The copy assigned to the control group $j_{ctrl,i}$ is censored when subject $i$ initiates treatment during the grace period, and the copy assigned to the treatment group $j_{trt,i}$ is censored if subject $i$ has not initiated treatment by the end of the grace period. (iii) Apply weighting to account for potential selection bias introduced by artificial censoring. With cloning, treatment and control groups have exactly the same distribution of baseline confounders, achieving perfect balance and "randomization" as in clinical trials. However, they have different censoring patterns, which cause selection bias. For example, those who are healthier and more likely to initiate treatment have treated copies followed until outcome but have control copies stop being followed when treatment is initiated. Those never-treated have control copies followed until outcome but have treated copies followed by the end of grace period, which is longer than all time to treatment initiation. Then, even when the true treatment effect is null, control copies still contribute less person-time than treatment copies. Therefore, weights adjusting for consequent selection bias caused by censoring are needed to generate unbiased estimation. Options include inverse probability treatment (IPT) weights (also called adherence weights) and inverse probability censoring (IPC) weights. IPT weights ensure that measured confounders are always balanced between the two groups over time, and IPC weights guarantee that censored and uncensored subjects within each group have the same distribution of measured confounders and thus ensure balanced measured confounders during follow-up.[39, 76]

For the exemplar, Steps (i) and (ii) are presented in Figure 6 (b). Here the grace period is taken as from study entry to the end of the 5th month. For subject 1, the treatment copy is followed until the observed censoring, and the control copy stops being followed at the end of the 4th month when treatment is initiated. For subject 2, the treatment copy is censored at the end of the 5th month, and the control copy is censored either at the observed censoring (i.e., an ITT analog) or later treatment initiation (i.e., a PP analog). For subject 4, the treatment copy is followed until outcome, and the control copy is artificially censored at the time of treatment initiation. For subjects 5, 6, and 7, the treatment copies are censored at the end of the grace period, and the control copies are followed until outcome. For subject 8, both copies are followed until the end of the 5$^{th}$ month. Similar graphical presentations are available in Petito et al. (2020)[76] and Maringe et al. (2020).[40]

For practical applications, Petito et al. (2020)[76] designed and emulated two targeted trials, aiming at evaluating the treatment effect of adjuvant fluorouracil initiation vs. no initiation within three months of curative surgery (tumor section) for individuals with stage II colorectal cancer as well as erlotinib initiation vs. no initiation within twelve weeks of gemcitabine initiation for individuals with advanced pancreatic adenocarcinoma. Cloning was adopted as the first step. For the first trial, after subjects met eligibility, time zero was set as the date of post-surgery discharge from hospital, and grace period was set as three months after that. For the second trial, time zero was the date of the first gemcitabine claim, and grace period was set as twelve weeks after that. The IPT weighting methods were subsequently adopted.

## 2.6 Software

Currently, to the best of our knowledge, there is no package specifically dedicated to for IMT analysis. However, packages or functions for some methods are publicly available and have been used in the field. Methods that model time zero can be realized by first creating a data frame containing defined time zero, treatment status, outcome, and covariate information for eligible subjects, and then adopting standard survival analysis functions.[77] The second family of methods can be more complicated. With SAS, R, SPSS, and some other software, the Cox regression procedures can be extended to incorporate time-dependent treatment indicator.[49, 78, 79] However, this may require a special data format and bring extra computational burden.[65] With R and SAS, marginal structural modeling can also be conducted to accommodate both time-varying treatment and time-varying confounders,[70, 80, 81] and the software can be found at https://www.hsph.harvard.edu/causal/software/. For Kaplan-Meier analysis with time-varying treatment indicator, Snapinn et al. (2005)[60] showed how to implement with software such as S-Plus and provided a graphical illustration. For the third family of methods, research codes have been provided by some researchers along with their publications. For example, Maringe et al. (2020)[40] provided sample R code for the cloning approach with IPC weights. Petito et al. (2020)[76] provided R code for the cloning approach with IPT weights.

## 3. Simulation

### 3.1 Settings

We resort to simulation to compare performance of the reviewed methods. With known data generating mechanisms, we can directly evaluate bias, which is not feasible with real-world data. We note that, although there has been some effort in the literature, comprehensive and direct comparison – as to be conducted here – is lacking.

We adopt the permutation algorithm, which has been widely used for generating survival times with prespecified distributions and conditional on baseline variables and time-varying treatment status.[82-84] Our procedure mostly follows that of Karim et al. (2016).[46] Specifically, for subject $i$ $(= 1, ..., n)$, the survival and censoring time pair $(T_{F,i}, T_{C,i})$ is generated from a pre-specified distribution. If $\min(T_{F,i}, T_{C,i}) > t_{K+1}$, $T_{C,i}$ is reset to $t_{K+1}$ (and subject $i$ is administratively censored). Accordingly, the censoring indicator $C_i = 1$ if $T_{F,i} \leq T_{C,i}$, and $C_i = 0$ otherwise. $(T_{F,i}, C_i)$'s are sorted with respect to $T_{F,i}$ in an increasing order and will be further matched with a covariate matrix.

We generate the covariate matrix $L_i$ with dimension $K \times N_L$, and the $k$th row corresponds to $L_{k,i} = (L_{0,i}, A_{k,i})$. We consider the case with only one binary covariate (beyond the indicator of time-varying treatment status). We acknowledge that this setting is overly simplified, however, increases our ability to appreciate the impact of covariate distributions and correlations and enable us to better focus on the treatment effect. Then $L_{0,i}$ is simply a scalar equal to 1 or 0, for example, and $N_L = 2$. Here $L_{0,i}$ is randomly generated from a Bernoulli distribution with probability of success 0.3. $A_{k,i}$ for subject $i$ and interval $k$ is generated based on the randomly assigned $T_{A,i}$. The treatment initiation time $T_{A,i}$ is generated from a uniform distribution $\text{Unif}(1, K)$, and a certain proportion of subjects do not initiate treatment during the entire follow-up regardless of outcome and censoring. Note that with this specific data generation, the binary $L_{0,i}$ is not a confounder (that influences both treatment initiation and outcome) but rather a covariate (that only influences outcome).

With this algorithm, a permutation probability law based on the Cox model partial likelihood is used as the basis for performing matching between $(T_{F,i}, C_i)$ and $L_i$. Specifically, at each sorted time $T_{F,i}$, $L_{i'}$ is sampled with probability $p_{i,k,L_{i'}}$ specified as follows:

$$\text{if } C_i = 1, p_{i,k,L_{i'}} = \frac{\exp(\beta_1 L_{0,i'} + \beta_2 A_{k,i'})}{\sum_{i'' \in r_i} \exp(\beta_1 L_{0,i''} + \beta_2 A_{k,i''})}; \text{ and if } C_i = 0, p_{i,k,L} = \frac{1}{\sum_{i'' \in r_i} 1}.$$

Here, $\beta_1$ is the log hazard ratio (HR) of the binary covariate, and $\beta_2$ is the log HR of the time-varying treatment; and $r_i$ denotes the at-risk set for this specific $(T_{F,i}, C_i)$ pair and contains subjects that have not been matched with previous survival pairs yet. Subjects that have already been matched with covariates will be removed from future matching.

Based on this algorithm, six scenarios are created, with different survival time distributions, proportions of never-treated subjects, and frequencies of events. For each scenario, we generate $B = 1,000$ datasets, sample size is set as $n = 5,000$, and the maximum observed time intervals is $K = 30$. For the binary covariate, we set log HR=-0.7; and for the time-varying treatment effect, log HR is set as 0.5. The six scenarios are summarized in Table 1.

|   | $T_{F,i} \sim$ | $T_{C,i} \sim$ | $T_{A,i} \sim U(1,30)$ |
|---|---|---|---|
| 1 | $\exp(\lambda = 0.01)$ | U(1,60) | 25% subjects have infinite $T_{A,i}$ |
| 2 | $\exp(\lambda = 0.01)$ | U(1,60) | 50% subjects have infinite $T_{A,i}$ |
| 3 | $\exp(\lambda = 0.01)$ | U(1,60) | 75% subjects have infinite $T_{A,i}$ |
| 4 | $\exp(\lambda = 0.1)$ | U(1,60) | 50% subjects have infinite $T_{A,i}$ |
| 5 | Gamma($\theta = 100, k = 0.4$) | U(1,60) | 50% subjects have infinite $T_{A,i}$ |
| 6 | Weibull($\eta = 100, k = 2$) | U(1,60) | 50% subjects have infinite $T_{A,i}$ |

**Table 1.** Simulation settings

To better illustrate, we generate data for each scenario with sample size $n = 50,000$ and present the Kaplan-Meier curves (Figure 7) for the treatment and control groups under the naïve definition of ever- and never-treated. Scenarios 1, 2, and 3 have the same survival distribution but different proportions of subjects who ever initiate treatment during follow-up. Scenario 2, 4, 5, and 6 have the same proportion of never-treated subjects but different survival distributions. Among them, scenario 5 may lead to the most severe bias if IMT is not properly handled, because of its highest risk of developing event early.

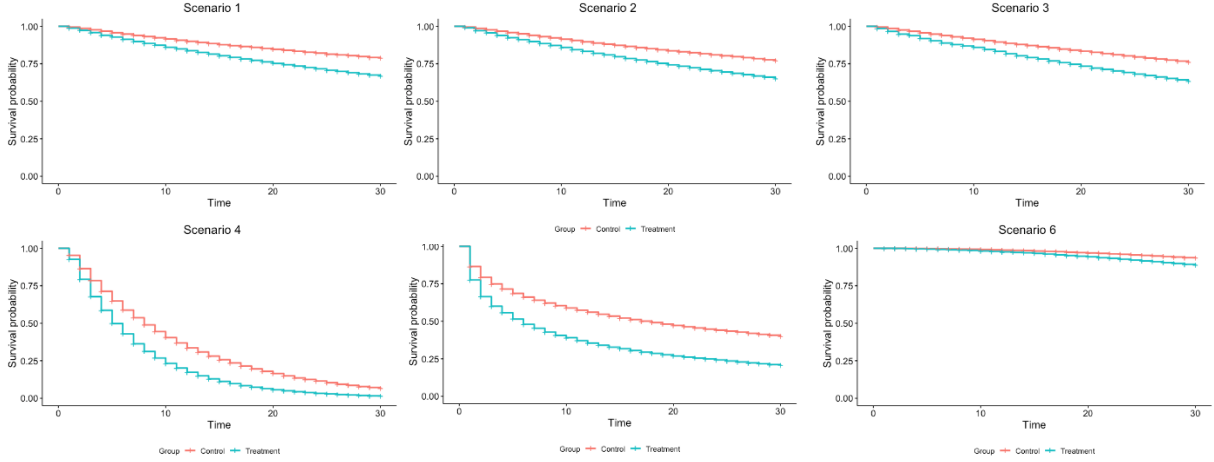

*Figure 7 Kaplan-Meier curves for the simulated data*

## 3.2 Evaluation metrics

We adopt the following metrics to evaluate performance: (1) Bias: $\sum_{b=1}^{B}(\hat{\beta}_{2,b} - \beta_2)/B$, (2) standard deviation (SD): $\sqrt{\sum_{b=1}^{B}\left(\hat{\beta}_{2,b} - \sum_{b=1}^{B}\frac{\hat{\beta}_{2,b}}{B}\right)^2 /(B-1)}$, (3) model-based (average) standard error (SE): average of the $B$ estimated standard errors of $\hat{\beta}_{2,b}$, which are directly obtained from software output, (4) coverage probability (CP) of the model-based 95% confidence intervals, and (5) mean square error (MSE): $\sqrt{\sum_{b=1}^{B}(\hat{\beta}_{2,b} - \beta_2)^2/B}$.

## 3.3 Results

All the aforementioned methods are implemented under the Cox model. We note that the additional binary variable is not a confounder but a covariate. As such, we simply add it to the model as opposed to adopting complex weighting or matching. The results are summarized in Table 2.

Across all the scenarios, the naïve classification of treatment groups (as ever- and never-treated) leads to negatively biased estimators, regardless of time zero definitions. Under Scenarios 1-3, among the first family of methods, including IMT leads to the highest absolute bias, while PTDM leads to the least. For example, for Scenario 1, the biases for including IMT, excluding IMT, PTDM are -1.387, -0.705, and -0.578, respectively. A higher proportion of subjects who initiate treatment during follow-up leads to a higher absolute bias, lower variance, and overall higher MSE. For example, for including IMT, the biases for Scenarios 1-3 are -1.387, -1.024, and -0.814, respectively, with

variances 0.063, 0.065, and 0.083, and MSEs 1.388, 1.026, and 0.818. Scenarios 2, 4, 5, and 6 have different event rates and distributions. It is observed that Scenario 5 has the most biases, which lead to the highest MSEs and lowest CPs. Under this scenario, the biases for including IMT, excluding IMT, and PTDM are -1.409, -0.933, and -0.417, respectively. As can be seen from Figure 7, the early risk of developing an event is the highest under this scenario. Under Scenario 6, bias is relatively small, but variance is relatively large, and estimation is unstable, because of the extremely low event rate. For determining the value of landmark, to be "less subjective", we consider the end of month 3, 6, and 9. It is observed that landmark analysis has much lower absolute biases and MSEs and much higher CPs across all scenarios. Different landmark choices lead to different performances for different scenarios, underlying the importance of this decision. Based on the six simulated scenarios, we observe that, when the rate of early event goes up (for example, scenarios 4 and 5 compared to scenarios 2 and 6), a bigger landmark value leads to more eligible samples excluded, resulting in more biased estimation and higher variance. When the treatment initiation rate is relatively low (for example, scenario 1 compared to scenario 2), a bigger landmark value may still not be able to catch enough treatment cases. It may help reduce variance but not bias, which is partly reflected in the low CP values. We observe that all the model-based SEs are close to SDs.

With the second family of methods, we observe that the time-varying modeling gives almost unbiased estimation, regardless of the proportion of subjects initiating treatment, overall event rate, and early risk of developing event. Variance is observed to have a higher dependence on the overall event rate and proportion of subjects with treatment initiation. For example., under Scenario 6 when the event rate is extremely low, though estimation is almost unbiased, a low coverage probability (73.8%) is observed, which is attributable to the high variance (0.127). When event rate is relatively high as under Scenarios 4 and 5, both SDs (0.045 and 0.056) and model-based SEs (0.044 and 0.056) are relatively low.

For the sequential trial method, we estimate PP treatment effects under the stratified Cox model weighted by IPC weights estimated from the Aalen's additive regression model.[43, 85, 86] This choice is made as the true treatment effect in our simulated data is PP. We find that the time-varying modeling and sequential trial methods both lead to almost unbiased estimation. Similar to the other methods, as event rate goes up and number of treatment initiations goes down, variance of estimates gets smaller. For example, for Scenarios 6, 2, 5, and 4 (which have increasing overall event rates), SDs are 0.136, 0.073, 0.057, and 0.045, respectively. We note that the robust sandwich estimator does account for multiple enrollments of the same subject in one analysis and leads to results similar to the standard deviations of estimated treatment effects. The CP values are close to 95% for all scenarios.

For the cloning method, we follow Maringe et al. (2020).[40] We note that in our simulation, treatment can initiate throughout the follow-up period with a relatively constant rate, which does not satisfy the assumption that a grace period exists and most of treatment initiation happens within the grace period. That is, there is no clearly defined and clinically relevant grace period in our case. To be comprehensive, we also consider multiple grace periods: 3 months, 6 months, and 9 months. In the original proposal by Maringe et al. (2020)[40], bootstrap was used for inference. As such, we do not provide model-based SE and CP results here. It is observed that all three grace period values lead to biased estimation, and the levels of bias are similar to those of PTDM. In addition, there is no optimal grace period choice. For example, for scenarios 1 to 5, a longer grace period results in a higher bias, lower variance, and overall higher MSE. For scenario 6, the reverse is true.

| Method | Scenario | Bias | SD | Model-based SE | CP | MSE |
|---|---|---|---|---|---|---|
| Include IMT | 1 | -1.387 | 0.063 | 0.065 | 0 | 1.388 |
|  | 2 | -1.024 | 0.065 | 0.070 | 0 | 1.026 |
|  | 3 | -0.814 | 0.083 | 0.087 | 0 | 0.818 |
|  | 4 | -0.972 | 0.040 | 0.043 | 0 | 0.973 |
|  | 5 | -1.409 | 0.048 | 0.052 | 0 | 1.409 |
|  | 6 | -0.654 | 0.121 | 0.121 | 0 | 0.665 |
| Exclude IMT | 1 | -0.705 | 0.062 | 0.065 | 0 | 0.708 |
|  | 2 | -0.368 | 0.071 | 0.071 | 0.1% | 0.375 |
|  | 3 | -0.132 | 0.085 | 0.088 | 68.5% | 0.157 |
|  | 4 | -0.061 | 0.048 | 0.043 | 70.7% | 0.078 |
|  | 5 | -0.933 | 0.051 | 0.052 | 0 | 0.935 |
|  | 6 | 0.283 | 0.121 | 0.124 | 37.1% | 0.307 |
| PTDM | 1 | -0.578 | 0.081 | 0.082 | 0 | 0.584 |
|  | 2 | -0.273 | 0.082 | 0.081 | 8.4% | 0.285 |
|  | 3 | -0.102 | 0.091 | 0.094 | 81.6% | 0.137 |
|  | 4 | -0.138 | 0.048 | 0.046 | 16.0% | 0.146 |
|  | 5 | -0.417 | 0.058 | 0.061 | 0 | 0.421 |
|  | 6 | -0.148 | 0.134 | 0.134 | 81.5% | 0.200 |
| Landmark Analysis (Landmark = 3) | 1 | -0.161 | 0.100 | 0.102 | 67.1% | 0.190 |
|  | 2 | -0.113 | 0.122 | 0.120 | 85.4% | 0.166 |
|  | 3 | -0.064 | 0.170 | 0.162 | 92.3% | 0.181 |
|  | 4 | -0.048 | 0.073 | 0.073 | 89.8% | 0.087 |
|  | 5 | -0.071 | 0.096 | 0.095 | 87.7% | 0.119 |
|  | 6 | -0.173 | 0.213 | 0.212 | 86.7% | 0.274 |
| Landmark Analysis | 1 | -0.161 | 0.089 | 0.090 | 58.1% | 0.184 |

| Method | | Bias | SD | SE | CP | MSE |
|---|---|---|---|---|---|---|
| (Landmark = 6) | 2 | -0.105 | 0.106 | 0.103 | 84.0% | 0.149 |
| | 3 | -0.058 | 0.143 | 0.136 | 91.1% | 0.154 |
| | 4 | -0.054 | 0.072 | 0.071 | 87.3% | 0.090 |
| | 5 | -0.078 | 0.094 | 0.088 | 83.8% | 0.122 |
| | 6 | -0.152 | 0.169 | 0.170 | 85.0% | 0.227 |
| Landmark Analysis | 1 | -0.160 | 0.086 | 0.089 | 56.6% | 0.182 |
| (Landmark = 9) | 2 | -0.099 | 0.104 | 0.099 | 82.2% | 0.144 |
| | 3 | -0.053 | 0.128 | 0.128 | 93.4% | 0.138 |
| | 4 | -0.056 | 0.077 | 0.076 | 88.8% | 0.095 |
| | 5 | -0.081 | 0.097 | 0.089 | 82.1% | 0.126 |
| | 6 | -0.128 | 0.153 | 0.153 | 86.4% | 0.199 |
| Time-varying Cox | 1 | -0.002 | 0.073 | 0.072 | 95.2% | 0.073 |
| | 2 | -0.001 | 0.074 | 0.075 | 94.4% | 0.074 |
| | 3 | -0.004 | 0.089 | 0.090 | 88.1% | 0.089 |
| | 4 | 0.000 | 0.045 | 0.044 | 99.7% | 0.045 |
| | 5 | -0.006 | 0.056 | 0.056 | 99% | 0.056 |
| | 6 | -0.003 | 0.127 | 0.126 | 73.8% | 0.127 |
| Sequential Cox | 1 | 0.006 | 0.079 | 0.079 | 94% | 0.080 |
| | 2 | -0.001 | 0.073 | 0.077 | 95.3% | 0.073 |
| | 3 | -0.004 | 0.091 | 0.090 | 94.7% | 0.091 |
| | 4 | -0.005 | 0.045 | 0.045 | 95.3% | 0.045 |
| | 5 | 0.007 | 0.057 | 0.058 | 95.7% | 0.057 |
| | 6 | 0.000 | 0.136 | 0.128 | 93.3% | 0.136 |
| Cloning Cox | 1 | -0.206 | 0.088 | - | - | 0.224 |
| (Grace period = 3) | 2 | -0.166 | 0.108 | - | - | 0.198 |
| | 3 | -0.132 | 0.149 | - | - | 0.199 |
| | 4 | -0.266 | 0.047 | - | - | 0.270 |
| | 5 | -0.314 | 0.047 | - | - | 0.317 |
| | 6 | -0.178 | 0.210 | - | - | 0.275 |
| Cloning Cox | 1 | -0.237 | 0.068 | - | - | 0.246 |
| (Grace period = 6) | 2 | -0.201 | 0.081 | - | - | 0.216 |
| | 3 | -0.177 | 0.109 | - | - | 0.208 |
| | 4 | -0.331 | 0.030 | - | - | 0.333 |
| | 5 | -0.341 | 0.037 | - | - | 0.343 |
| | 6 | -0.167 | 0.161 | - | - | 0.232 |
| Cloning Cox | 1 | -0.261 | 0.056 | - | - | 0.267 |
| (Grace period = 9) | 2 | -0.233 | 0.068 | - | - | 0.243 |
| | 3 | -0.218 | 0.083 | - | - | 0.233 |
| | 4 | -0.368 | 0.022 | - | - | 0.368 |
| | 5 | -0.365 | 0.030 | - | - | 0.366 |
| | 6 | -0.161 | 0.137 | - | - | 0.212 |

**Table 2**. Simulation results. In each cell, mean based on 1,000 replicates.

Summarily, simulation suggests that the time-varying modeling and sequential trial methods lead to almost unbiased estimation. The time-varying modeling method has a smaller variance than the sequential trial method. The methods with time-independent treatment status and redefined time zero can partly account for IMT but still lead to biased estimation. The amount of bias and variance depend on event rate and distribution of time to treatment initiation. In our simulation, all methods using redefined time zero have negative bias, with including IMT having the most severe bias and landmark analysis having the least.

## 4. Analysis of the Stanford heart transplant data

### 4.1 Data

We re-analyze the seminal Stanford heart transplant data. The goal is not to make new scientific discoveries – this data has been thoroughly analyzed, and its limitations have been well recognized. Rather, the goal is to demonstrate different performance of the review methods under a real-world setting with a "benchmark" dataset. The data is publicly available at https://www.openintro.org/data/index.php?data=heart_transplant. The background of this study has been fully described elsewhere. Briefly, this was a prospective cohort study aiming at evaluating cardiac transplantation for severely illed patients with end-stage heart disease. All patients were clinically designated as heart transplant candidates, and the waiting time ranged between a few weeks to several months depending on donor availability. A few candidates showed improvement and were deselected as heart transplant candidates during the waiting period. There were 103 subjects, among whom 69 received transplantation, and 34 either died, were censored, or were deselected before donors were available. Baseline variables include patient ID, year of acceptance as a heart transplant candidate, patient age at the beginning of the study, and prior surgery status. Follow-up and outcome variables include survival time, event indicator, transplantation indicator, and waiting time for ever-treated.

### 4.2 Implementation

The reviewed methods are implemented in the same manner as in the simulation. As a clinically relevant landmark is lacking, as in simulation, we consider multiple values including 20, 40, and 60 days. When implementing the sequential method, we observe that most of the treatment initiations happened in the early stage of follow-up (the 1st quartile of waiting days is 10, and the 3rd quartile is 46). We consider three subsequent sub-trials, each with an enrollment window of 20 days. The first day of each sub-trial is defined as time zero. For the cloning approach, we also consider multiple grace periods: 20, 40, and 60 days. With all methods, the Cox model is adopted to evaluate the effect of heart transplantation on time to death. Age and prior surgery status are included as covariates.[12, 87, 88] For the three variables, HRs, 95% confidence intervals, and p-values are obtained. It is noted that with the cloning method, the effects of covariates are incorporated and evaluated from the weighting process. Therefore, estimates for the covariates are not available.

### 4.3 Results

Table 3 shows that, similar to in simulation, different methods lead to significantly different results. In particular, the including IMT, excluding IMT, and PTDM methods lead to significant estimated HRs that are less than 1. The other methods lead to non-significant HRs. In landmark analysis, a larger landmark value leads to a smaller HR. The time-varying Cox and sequential methods generate estimated HRs 0.92 and 0.81, respectively. With the cloning method, a longer grade period leads to a smaller HR. The analysis results with the second and third families of methods align with many published studies. For example, Turnbull et al. (1974)[11] adopted three non-parametric and three parametric methods and concluded that all models suggested small, non-significant beneficial treatment effects.

| Method | Variable | HR | 95% CI | P-value |
|---|---|---|---|---|
| Include IMT | transplant (ever received) | 0.19 | 0.11, 0.33 | < 0.0001 |
| | age | 1.06 | 1.03, 1.09 | < 0.0001 |
| | prior surgery (yes) | 0.48 | 0.20, 1.13 | 0.093 |
| Exclude IMT | transplant (ever received) | 0.25 | 0.15, 0.42 | < 0.0001 |
| | age | 1.06 | 1.03, 1.09 | 0.0002 |
| | prior surgery (yes) | 0.48 | 0.20, 1.14 | 0.097 |
| PTDM | transplant (ever received) | 0.22 | 0.12, 0.42 | < 0.0001 |
| | age | 1.06 | 1.02, 1.09 | 0.0006 |
| | prior surgery (yes) | 0.49 | 0.21, 1.17 | 0.107 |
| Landmark Analysis (Landmark = 20) | transplant (ever received) | 1.36 | 0.77, 2.42 | 0.292 |
| | age | 1.04 | 1.00, 1.07 | 0.025 |
| | prior surgery (yes) | 0.41 | 0.17, 0.96 | 0.041 |
| Landmark Analysis (Landmark = 40) | transplant (ever received) | 0.91 | 0.50, 1.65 | 0.754 |
| | age | 1.05 | 1.01, 1.09 | 0.017 |
| | prior surgery (yes) | 0.43 | 0.18, 1.04 | 0.061 |
| Landmark Analysis (Landmark = 60) | transplant (ever received) | 0.62 | 0.30, 1.28 | 0.193 |
| | age | 1.07 | 1.02, 1.13 | 0.003 |
| | prior surgery (yes) | 0.52 | 0.22, 1.25 | 0.145 |
| Time-varying Cox | transplant (ever received) | 0.92 | 0.51, 1.66 | 0.773 |
| | age | 1.03 | 1.01, 1.06 | 0.017 |
| | prior surgery (yes) | 0.33 | 0.14, 0.77 | 0.010 |
| Sequential Cox | transplant (ever received) | 0.81 | 0.57, 1.16 | 0.249 |
|  | age | 1.03 | 1.00, 1.07 | 0.053 |
|  | prior surgery (yes) | 0.32 | 0.15, 0.67 | 0.003 |
| Cloning Cox (Grace period = 20) | transplant (ever received) | 1.31 | 0.90, 2.06 | - |
| Cloning Cox (Grace period = 40) | transplant (ever received) | 0.99 | 0.71, 1.41 | - |
| Cloning Cox (Grace period = 60) | transplant (ever received) | 0.84 | 0.59, 1.17 | - |

HR *hazards ratio*, CI *confidence interval*

**Table 3.** Analysis results for the Stanford heart transplant data

## 5. Connections to causal inference analysis

In epidemiologic studies, IMT bias often interconnects with other bias issues, in particular confounding bias in causal inference analysis, leading to more challenges. Causal inference is a large field, and we refer to Hernan and Robins (2020) [89], Morgan and Winship (2015) [90] and others for relevant discussions. Our literature review suggests that there has been insufficient attention to the methods reviewed above when a causal effect is of interest. In Appendix, we address this knowledge gap and discuss the possibility of combining the aforementioned methods with matching, weighting, and emulation approaches under the causal inference paradigm.

## 6. Remarks

In this article, we have conducted a selected review of methods for accommodating IMT. This review may assist biomedical researchers to better understand and address IMT in the analysis of observational studies and clinical trials. We have focused on intuition, rationale, and numerical performance, without delving into mathematical details. Interested readers are referred to the original publications for the technical aspects. We acknowledge that some relevant approaches and applications may have inevitably been overlooked. It is also recognized that accommodating IMT is still a developing area of research, and that this review may need to be updated in the future.

**Funding Acknowledgements**

This research received no specific grant from any funding agency in the public, commercial, or not-for-profit sectors.

**Declaration of Conflicting Interests**

The Authors declare that there is no conflict of interest.

**References**

1. Buyse M and Piedbois P. On the relationship between response to treatment and survival time. *Statistics in Medicine* 1996; 15: 2797-2812.
2. Peduzzi P, Detre K, Wittes J, et al. Intent-to-treat analysis and the problem of crossovers. An example from the Veterans Administration coronary bypass surgery study. *J Thorac Cardiovasc Surg* 1991; 101: 481-487. 1991/03/01.
3. Austin PC, Mamdani MM, van Walraven C, et al. Quantifying the impact of survivor treatment bias in observational studies. *J Eval Clin Pract* 2006; 12: 601-612. 2006/11/15. DOI: 10.1111/j.1365-2753.2005.00624.x.
4. van Walraven C, Davis D, Forster AJ, et al. Time-dependent bias was common in survival analyses published in leading clinical journals. *Journal of Clinical Epidemiology* 2004; 57: 672-682. DOI: https://doi.org/10.1016/j.jclinepi.2003.12.008.
5. Ho AM, Dion PW, Yeung JH, et al. Simulation of survivorship bias in observational studies on plasma to red blood cell ratios in massive transfusion for trauma. *Br J Surg* 2012; 99 Suppl 1: 132-139. 2012/03/28. DOI: 10.1002/bjs.7732.
6. Delgado-Rodríguez M and Llorca J. Bias. *J Epidemiol Community Health* 2004; 58: 635-641. 2004/07/15. DOI: 10.1136/jech.2003.008466.
7. Walker AM. *Observation and inference: an introduction to the methods of epidemiology*. Newton Lower Falls, MA: Epidemiology Resources Inc., 1991.
8. Suissa S. Immortal time bias in pharmaco-epidemiology. *Am J Epidemiol* 2008; 167: 492-499. 2007/12/07. DOI: 10.1093/aje/kwm324.
9. Gail MH. Does cardiac transplantation prolong life? A reassessment. *Ann Intern Med* 1972; 76: 815-817. 1972/05/01. DOI: 10.7326/0003-4819-76-5-815.
10. Mantel N and Byar DP. Evaluation of Response-Time Data Involving Transient States: An Illustration Using Heart-Transplant Data. *Journal of the American Statistical Association* 1974; 69: 81-86. DOI: 10.1080/01621459.1974.10480131.
11. Turnbull BW, Brown BW and Hu M. Survivorship Analysis of Heart Transplant Data. *Journal of the American Statistical Association* 1974; 69: 74-80. DOI: 10.1080/01621459.1974.10480130.
12. Crowley J and Hu M. Covariance Analysis of Heart Transplant Survival Data. *Journal of the American Statistical Association* 1977; 72: 27-36. DOI: 10.1080/01621459.1977.10479903.
13. Suissa S. Effectiveness of inhaled corticosteroids in chronic obstructive pulmonary disease: immortal time bias in observational studies. *Am J Respir Crit Care Med* 2003; 168: 49-53. 2003/03/29. DOI: 10.1164/rccm.200210-1231OC.
14. Suissa S. Inhaled steroids and mortality in COPD: bias from unaccounted immortal time. *Eur Respir J* 2004; 23: 391-395. 2004/04/07. DOI: 10.1183/09031936.04.00062504.
15. Sin DD and Tu JV. Inhaled corticosteroids and the risk of mortality and readmission in elderly patients with chronic obstructive pulmonary disease. *Am J Respir Crit Care Med* 2001; 164: 580-584. 2001/08/25. DOI: 10.1164/ajrccm.164.4.2009033.
16. Samet JM. Measuring the effectiveness of inhaled corticosteroids for COPD is not easy! *Am J Respir Crit Care Med* 2003; 168: 1-2. 2003/06/27. DOI: 10.1164/rccm.2304004.
17. Soriano JB, Vestbo J, Pride NB, et al. Survival in COPD patients after regular use of fluticasone propionate and salmeterol in general practice. *Eur Respir J* 2002; 20: 819-825. 2002/11/05. DOI: 10.1183/09031936.02.00301302.
18. Sin DD and Tu JV. Inhaled corticosteroid therapy reduces the risk of rehospitalization and all-cause mortality in elderly asthmatics. *Eur Respir J* 2001; 17: 380-385. 2001/06/19. DOI: 10.1183/09031936.01.17303800.
19. Adams RJ, Fuhlbrigge AL, Finkelstein JA, et al. Intranasal steroids and the risk of emergency department visits for asthma. *J Allergy Clin Immunol* 2002; 109: 636-642. 2002/04/10. DOI: 10.1067/mai.2002.123237.
20. Glesby MJ and Hoover DR. Survivor treatment selection bias in observational studies: examples from the AIDS literature. *Ann Intern Med* 1996; 124: 999-1005. 1996/06/01. DOI: 10.7326/0003-4819-124-11-199606010-00008.

21. Mi X, Hammill BG, Curtis LH, et al. Impact of immortal person-time and time scale in comparative effectiveness research for medical devices: a case for implantable cardioverter-defibrillators. *J Clin Epidemiol* 2013; 66: S138-144. 2013/07/17. DOI: 10.1016/j.jclinepi.2013.01.014.
22. Tseng VL, Chlebowski RT, Yu F, et al. Association of Cataract Surgery With Mortality in Older Women: Findings from the Women's Health Initiative. *JAMA Ophthalmology* 2018; 136: 3-10. DOI: 10.1001/jamaophthalmol.2017.4512.
23. Harding BN and Weiss NS. Point: Immortal Time Bias-What Are the Determinants of Its Magnitude? *Am J Epidemiol* 2019; 188: 1013-1015. 2019/06/04. DOI: 10.1093/aje/kwz067.
24. Bernasconi DP, Rebora P, Iacobelli S, et al. Survival probabilities with time-dependent treatment indicator: quantities and non-parametric estimators. *Stat Med* 2016; 35: 1032-1048. 2015/10/28. DOI: 10.1002/sim.6765.
25. Gray R and Wheatley K. How to avoid bias when comparing bone marrow transplantation with chemotherapy. *Bone Marrow Transplant* 1991; 7 Suppl 3: 9-12. 1991/01/01.
26. Duck BW, Carter JT and Coombes EJ. Mortality study of workers in a polyvinyl-chloride production plant. *Lancet* 1975; 2: 1197-1199. 1975/12/13. DOI: 10.1016/s0140-6736(75)92674-4.
27. Enterline PE. Pitfalls in epidemiological research. An examination of the asbestos literature. *J Occup Med* 1976; 18: 150-156. 1976/03/01.
28. Wagoner JK, Infante PF and Saracci R. Vinyl chloride and mortality? *Lancet* 1976; 2: 194-195. 1976/07/24. DOI: 10.1016/s0140-6736(76)92361-8.
29. May GS, DeMets DL, Friedman LM, et al. The randomized clinical trial: bias in analysis. *Circulation* 1981; 64: 669-673. 1981/10/01. DOI: 10.1161/01.cir.64.4.669.
30. Hernán MA and Robins JM. Per-Protocol Analyses of Pragmatic Trials. *N Engl J Med* 2017; 377: 1391-1398. 2017/10/05. DOI: 10.1056/NEJMsm1605385.
31. Sheiner LB and Rubin DB. Intention-to-treat analysis and the goals of clinical trials. *Clin Pharmacol Ther* 1995; 57: 6-15. 1995/01/01. DOI: 10.1016/0009-9236(95)90260-0.
32. Gupta SK. Intention-to-treat concept: A review. *Perspect Clin Res* 2011; 2: 109-112. 2011/09/08. DOI: 10.4103/2229-3485.83221.
33. Murray EJ, Caniglia EC and Petito LC. Causal survival analysis: A guide to estimating intention-to-treat and per-protocol effects from randomized clinical trials with non-adherence. *Research Methods in Medicine & Health Sciences* 2021; 2: 39-49.
34. Peduzzi P, Wittes J, Detre K, et al. Analysis as-randomized and the problem of non-adherence: an example from the Veterans Affairs Randomized Trial of Coronary Artery Bypass Surgery. *Stat Med* 1993; 12: 1185-1195. 1993/07/15. DOI: 10.1002/sim.4780121302.
35. Group VCABSCS. Eighteen-year follow-up in the Veterans Affairs Cooperative Study of Coronary Artery Bypass Surgery for stable angina. *Circulation* 1992; 86: 121-130. 1992/07/01. DOI: 10.1161/01.cir.86.1.121.
36. Anderson JR, Cain KC and Gelber RD. Analysis of survival by tumor response. *J Clin Oncol* 1983; 1: 710-719. 1983/11/01. DOI: 10.1200/jco.1983.1.11.710.
37. Zhou Z, Rahme E, Abrahamowicz M, et al. Survival bias associated with time-to-treatment initiation in drug effectiveness evaluation: a comparison of methods. *Am J Epidemiol* 2005; 162: 1016-1023. 2005/09/30. DOI: 10.1093/aje/kwi307.
38. Hernán MA, Sauer BC, Hernández-Díaz S, et al. Specifying a target trial prevents immortal time bias and other self-inflicted injuries in observational analyses. *Journal of Clinical Epidemiology* 2016; 79: 70-75. DOI: https://doi.org/10.1016/j.jclinepi.2016.04.014.
39. Hernán MA. How to estimate the effect of treatment duration on survival outcomes using observational data. *BMJ* 2018; 360: k182. DOI: 10.1136/bmj.k182.
40. Maringe C, Benitez Majano S, Exarchakou A, et al. Reflection on modern methods: trial emulation in the presence of immortal-time bias. Assessing the benefit of major surgery for elderly lung cancer patients using observational data. *Int J Epidemiol* 2020; 49: 1719-1729. 2020/05/10. DOI: 10.1093/ije/dyaa057.
41. Thomas LE, Yang S, Wojdyla D, et al. Matching with time-dependent treatments: A review and look forward. *Statistics in Medicine* 2020; 39: 2350-2370. DOI: https://doi.org/10.1002/sim.8533.
42. Simon R and Makuch RW. A non-parametric graphical representation of the relationship between survival and the occurrence of an event: application to responder versus non-responder bias. *Stat Med* 1984; 3: 35-44. 1984/01/01. DOI: 10.1002/sim.4780030106.
43. Gran JM, Røysland K, Wolbers M, et al. A sequential Cox approach for estimating the causal effect of treatment in the presence of time-dependent confounding applied to data from the Swiss HIV Cohort Study. *Stat Med* 2010; 29: 2757-2768. 2010/08/31. DOI: 10.1002/sim.4048.

44. Yadav K and Lewis RJ. Immortal Time Bias in Observational Studies. *Jama* 2021; 325: 686-687. 2021/02/17. DOI: 10.1001/jama.2020.9151.
45. Platt RW, Hutcheon JA and Suissa S. Immortal Time Bias in Epidemiology. *Current Epidemiology Reports* 2019; 6: 23-27. DOI: 10.1007/s40471-019-0180-5.
46. Karim ME, Gustafson P, Petkau J, et al. Comparison of Statistical Approaches for Dealing With Immortal Time Bias in Drug Effectiveness Studies. *Am J Epidemiol* 2016; 184: 325-335. 2016/07/28. DOI: 10.1093/aje/kwv445.
47. Mi X, Hammill BG, Curtis LH, et al. Use of the landmark method to address immortal person-time bias in comparative effectiveness research: a simulation study. *Stat Med* 2016; 35: 4824-4836. 2016/06/29. DOI: 10.1002/sim.7019.
48. Sin DD and McAlister FA. The effects of beta-blockers on morbidity and mortality in a population-based cohort of 11,942 elderly patients with heart failure. *Am J Med* 2002; 113: 650-656. 2002/12/31. DOI: 10.1016/s0002-9343(02)01346-3.
49. Gleiss A, Oberbauer R and Heinze G. An unjustified benefit: immortal time bias in the analysis of time-dependent events. *Transpl Int* 2018; 31: 125-130. 2017/10/13. DOI: 10.1111/tri.13081.
50. Liu J, Weinhandl ED, Gilbertson DT, et al. Issues regarding 'immortal time' in the analysis of the treatment effects in observational studies. *Kidney Int* 2012; 81: 341-350. 2011/11/18. DOI: 10.1038/ki.2011.388.
51. Smith PK, Califf RM, Tuttle RH, et al. Selection of surgical or percutaneous coronary intervention provides differential longevity benefit. *Ann Thorac Surg* 2006; 82: 1420-1428; discussion 1428-1429. 2006/09/26. DOI: 10.1016/j.athoracsur.2006.04.044.
52. Mamdani M, Rochon PA, Juurlink DN, et al. Observational study of upper gastrointestinal haemorrhage in elderly patients given selective cyclo-oxygenase-2 inhibitors or conventional non-steroidal anti-inflammatory drugs. *Bmj* 2002; 325: 624. 2002/09/21. DOI: 10.1136/bmj.325.7365.624.
53. Keyhan G, Chen S-F and Pilote L. Angiotensin-converting enzyme inhibitors and survival in women and men with heart failure. *European journal of heart failure* 2007; 9: 594-601. DOI: 10.1016/j.ejheart.2007.03.004.
54. Pittet D, Tarara D and Wenzel RP. Nosocomial Bloodstream Infection in Critically III Patients: Excess Length of Stay, Extra Costs, and Attributable Mortality. *JAMA* 1994; 271: 1598-1601. DOI: 10.1001/jama.1994.03510440058033.
55. Guyatt GH, Sackett DL and Cook DJ. Users' guides to the medical literature. II. How to use an article about therapy or prevention. A. Are the results of the study valid? Evidence-Based Medicine Working Group. *Jama* 1993; 270: 2598-2601. 1993/12/01. DOI: 10.1001/jama.270.21.2598.
56. Fosså SD and Skovlund E. Selection of patients may limit the generalizability of results from cancer trials. *Acta Oncol* 2002; 41: 131-137. 2002/07/10. DOI: 10.1080/028418602753669490.
57. Putter H and van Houwelingen HC. Understanding Landmarking and Its Relation with Time-Dependent Cox Regression. *Stat Biosci* 2017; 9: 489-503. 2017/12/12. DOI: 10.1007/s12561-016-9157-9.
58. Herzog CA, Li S, Weinhandl ED, et al. Survival of dialysis patients after cardiac arrest and the impact of implantable cardioverter defibrillators. *Kidney Int* 2005; 68: 818-825. 2005/07/15. DOI: 10.1111/j.1523-1755.2005.00462.x.
59. Haller MC, Kammer M, Kainz A, et al. Steroid withdrawal after renal transplantation: a retrospective cohort study. *BMC Med* 2017; 15: 8. 2017/01/13. DOI: 10.1186/s12916-016-0772-6.
60. Snapinn SM, Jiang Q and Iglewicz B. Illustrating the Impact of a Time-Varying Covariate With an Extended Kaplan-Meier Estimator. *The American Statistician* 2005; 59: 301-307. DOI: 10.1198/000313005X70371.
61. Cox DR. Regression Models and Life-Tables. *Journal of the Royal Statistical Society: Series B (Methodological)* 1972; 34: 187-202. https://doi.org/10.1111/j.2517-6161.1972.tb00899.x. DOI: https://doi.org/10.1111/j.2517-6161.1972.tb00899.x.
62. Suissa S. Immortal time bias in observational studies of drug effects. *Pharmacoepidemiol Drug Saf* 2007; 16: 241-249. 2007/01/26. DOI: 10.1002/pds.1357.
63. Shintani AK, Girard TD, Eden SK, et al. Immortal time bias in critical care research: application of time-varying Cox regression for observational cohort studies. *Crit Care Med* 2009; 37: 2939-2945. 2009/09/23. DOI: 10.1097/CCM.0b013e3181b7fbbb.
64. Dekker FW, de Mutsert R, van Dijk PC, et al. Survival analysis: time-dependent effects and time-varying risk factors. *Kidney International* 2008; 74: 994-997. DOI: https://doi.org/10.1038/ki.2008.328.
65. Therneau T, Crowson C and Atkinson E. Using time dependent covariates and time dependent coefficients in the cox model. *Survival Vignettes* 2017; 2: 3.
66. Fisher LD and Lin DY. Time-dependent covariates in the Cox proportional-hazards regression model. *Annual review of public health* 1999; 20: 145-157.

67. Leffondré K, Abrahamowicz M and Siemiatycki J. Evaluation of Cox's model and logistic regression for matched case‐control data with time‐dependent covariates: a simulation study. *Statistics in medicine* 2003; 22: 3781-3794.
68. Therneau TM and Grambsch PM. The cox model. *Modeling survival data: extending the Cox model*. Springer, 2000, pp.39-77.
69. Mehrotra DV, Su SC and Li X. An efficient alternative to the stratified cox model analysis. *Statistics in medicine* 2012; 31: 1849-1856.
70. Hernán MA, Brumback B and Robins JM. Marginal structural models to estimate the causal effect of zidovudine on the survival of HIV-positive men. *Epidemiology* 2000; 11: 561-570. 2000/08/24. DOI: 10.1097/00001648-200009000-00012.
71. Danaei G, Garcia Rodriguez LA, Cantero OF, et al. Electronic medical records can be used to emulate target trials of sustained treatment strategies. *J Clin Epidemiol* 2018; 96: 12-22. 2017/12/06. DOI: 10.1016/j.jclinepi.2017.11.021.
72. Hardin JW. Generalized Estimating Equations (GEE). *Encyclopedia of Statistics in Behavioral Science*. 2005.
73. Hernán MA, Lanoy E, Costagliola D, et al. Comparison of dynamic treatment regimes via inverse probability weighting. *Basic Clin Pharmacol Toxicol* 2006; 98: 237-242. 2006/04/14. DOI: 10.1111/j.1742-7843.2006.pto_329.x.
74. Danaei G, Rodriguez LA, Cantero OF, et al. Observational data for comparative effectiveness research: an emulation of randomised trials of statins and primary prevention of coronary heart disease. *Stat Methods Med Res* 2013; 22: 70-96. 2011/10/22. DOI: 10.1177/0962280211403603.
75. Gupta S, Wang W, Hayek SS, et al. Association Between Early Treatment With Tocilizumab and Mortality Among Critically Ill Patients With COVID-19. *JAMA Intern Med* 2021; 181: 41-51. 2020/10/21. DOI: 10.1001/jamainternmed.2020.6252.
76. Petito LC, García-Albéniz X, Logan RW, et al. Estimates of Overall Survival in Patients With Cancer Receiving Different Treatment Regimens: Emulating Hypothetical Target Trials in the Surveillance, Epidemiology, and End Results (SEER)-Medicare Linked Database. *JAMA Netw Open* 2020; 3: e200452. 2020/03/07. DOI: 10.1001/jamanetworkopen.2020.0452.
77. Clark TG, Bradburn MJ, Love SB, et al. Survival Analysis Part I: Basic concepts and first analyses. *British Journal of Cancer* 2003; 89: 232-238. DOI: 10.1038/sj.bjc.6601118.
78. Altman DG and De Stavola BL. Practical problems in fitting a proportional hazards model to data with udated measurements of the covariates. *Statistics in medicine* 1994; 13: 301-341.
79. Clark TG, Bradburn MJ, Love SB, et al. Survival Analysis Part IV: Further concepts and methods in survival analysis. *British Journal of Cancer* 2003; 89: 781-786. DOI: 10.1038/sj.bjc.6601117.
80. Robins JM. Marginal Structural Models versus Structural nested Models as Tools for Causal inference. In: *Statistical Models in Epidemiology, the Environment, and Clinical Trials* (eds Halloran ME and Berry D), New York, NY, 2000, pp.95-133. Springer New York.
81. Robins JM, Hernán MA and Brumback B. Marginal structural models and causal inference in epidemiology. *Epidemiology* 2000; 11: 550-560. 2000/08/24. DOI: 10.1097/00001648-200009000-00011.
82. Abrahamowicz M, Mackenzie T and Esdaile JM. Time-Dependent Hazard Ratio: Modeling and Hypothesis Testing with Application in Lupus Nephritis. *Journal of the American Statistical Association* 1996; 91: 1432-1439. DOI: 10.1080/01621459.1996.10476711.
83. Mackenzie T and Abrahamowicz M. Marginal and hazard ratio specific random data generation: Applications to semi-parametric bootstrapping. *Statistics and Computing* 2002; 12: 245-252. DOI: 10.1023/A:1020750810409.
84. Sylvestre M-P and Abrahamowicz M. Comparison of algorithms to generate event times conditional on time-dependent covariates. *Statistics in Medicine* 2008; 27: 2618-2634. DOI: https://doi.org/10.1002/sim.3092.
85. Aalen O. A Model for Nonparametric Regression Analysis of Counting Processes. In: New York, NY, 1980, pp.1-25. Springer New York.
86. Aalen OO. A linear regression model for the analysis of life times. *Statistics in Medicine* 1989; 8: 907-925. DOI: https://doi.org/10.1002/sim.4780080803.
87. Aitkin M, Laird N and Francis B. A reanalysis of the Stanford heart transplant data. *Journal of the American Statistical Association* 1983; 78: 264-274.
88. Tsujitani M and Tanaka Y. Analysis of heart transplant survival data using generalized additive models. *Computational and mathematical methods in medicine* 2013; 2013.
89. Hernán MA and Robins JM. *Causal Inference: What If*. Boca Raton: Chapman & Hall/CRC, 2020.
90. Morgan SL and Winship C. *Counterfactuals and causal inference*. Cambridge University Press, 2015.

**Appendix for "Statistical Methods for Accommodating Immortal Time: A Selective Review and Comparison"**

**by Jiping Wang, Peter Peduzzi, Michael Wininger, Shuangge Ma**

**Appendix A: Accommodating IMT in causal inference analysis**

**A.1 Causal inference methods**

Causal inference is a large field. Here we only briefly mention causal inference methods that are relevant to the review of IMT bias correction methods. Weighting and matching have been broadly used in causal inference to alleviate confounding bias and selection bias which commonly exist in observational studies. Weighting and matching are usually performed based on quantities such as propensity scores calculated using covariates and treatment status.[1, 2] Causal inference based on observational data requires special techniques when treatment initiation happens during follow-up, and/or confounder values change over time and potentially inform future treatment. In this case, weighting and matching can be problematic, and more complex methods such as marginal structural models with time-varying weights are needed.[3-5]

Compared to weighting and matching, emulation is relatively more recent.[6] Under emulation, "clinical trials" are created using observational data, and statistical techniques for randomized clinical trials are then directly applied. Emulation requires mimicking randomization, and this is usually achieved with IPT weighting and propensity score matching. As such, emulation suffers from the same restriction of choosing methods caused by applying weighting and matching in analysis.[6-10] Emulation may not be feasible if the method selected for dealing with IMT is "backward" (that is, uses information that is only available after baseline to decide treatment status) and under time-dependent bias (i.e., bias caused by using "baseline immeasurable" information).[11] In addition, certain inconsistent components (e.g., time zero and follow-up) for treatment and control groups deviate from the concept of "randomization" and lose comparability in real and emulated target trials.

**A.2 Methods that model time zero**

These methods have been reviewed in Section 2.3 of the main text. As previously mentioned, PTDM and landmark analysis exclude some eligible subjects at study entry, while others do not. For causal inference analysis, this exclusion brings additional complication, as it distorts target population and thus impacts both internal and external validity of the estimated causal effect or even association.[12, 13] Therefore, estimation with certain exclusions should be interpreted with caution. For example, landmark analysis excludes those who die before redefined time zero. The subsequent estimation result is only applicable to those who are generally healthier and live longer.

When emulation is adopted, another layer of ambiguity appears. Under emulation, researchers analyze data in a way that they normally do with a clinical trial. This involves screening for eligibility, measuring baseline characteristics, assigning treatment groups before the start of follow-up, and then following subjects prospectively. With the reviewed methods, classification of treatment and control groups is performed based on data at baseline or during follow-up. If we combine those methods with emulation, we may need to question the validity of analyzing data in a backward direction. As such, it does not align with the principle of emulation. It has more of a PP flavor, as treatment status is determined based on what subjects "adhere" to in reality. In comparison, forward analysis assigns treatment groups at baseline regardless of future treatment switching and non-adherence, thus having more of an ITT flavor. Recall that the method of redefining time zero after grace period decides treatment status based on what happens during grace period but before redefined time zero. Similarly, landmark analysis assigns treatment groups based on what happens before landmark. These forward methods align well with the principle of emulation. After determining treatment status at time zero, one can accommodate differences in (distributions of) baseline covariates using weighting and matching. In addition to the forward concerns and generalizability, without the assumptions required for a valid grace period, landmark analysis is more likely to be descriptive than causal inference in the sense that it generates results that are conditional on observed IMT. This precludes a counterfactual interpretation that is required by causal inference.[14]

Another concern rises from the inconsistent definitions of time zero for treatment and control groups. In an actual clinical trial, time zero is often the time of randomization and treatment assignment, and time zero for both groups is either the same or set using the same logic. With observational data, "true" time zero is usually unknown and needs to be specified. For the excluding IMT method, time zero for the treatment group is time of treatment initiation, and time zero for the control group is study entry. These two definitions are incomparable, leading to an inappropriate emulation analysis.

**A.3 Methods that model time-varying treatment status**

Many causal inference methods involve weighting and matching to accommodate potential confounding and selection bias in observational studies and non-adherence in clinical trials.[15-19] A drawback of time-varying treatment modeling is that the time-varying definition of treatment status makes propensity score estimation much harder as it becomes time-dependent. In contrast, an "unexpected benefit" is that time-varying propensity scores and corresponding weights provide a better carrier of information on time-dependent confounders and prognostic factors. To simplify discussion, we have mostly focused on the scenario with time-independent covariates. Under such a scenario, the probability of treatment initiation at a specific time point may only depend on time and previous treatment status. If confounder values are time-dependent, time-varying propensity scores depend on treatment history and time-dependent confounders,[20] and marginal structural model approach is needed, which has time-varying propensity score weighting and can avoid time-varying confounding bias.[3-5]

With emulation, one concern is that it is not possible to estimate an ITT effect under the time-varying treatment definition because an ITT approach "holds" subjects in groups they belong to at baseline. However, at the beginning of follow-up, nearly all subjects have not initiated treatment and, hence, are classified as controls. However, if the adherence-adjusted effect is of interest and the aforementioned confounding problem is neglected, time-varying treatment status modeling is potentially appropriate. In some clinical trials, if covariate balance can be assumed at all time, the time-varying treatment definition can be adopted to help derive adherence-adjusted effect.[21-23] Another concern is that time-varying treatment status may conflict with the counterfactual interpretation that is required in causal inference.

**A.4 Sequential trial and cloning methods**

For the sequential trial method, matching or weighting is needed to address the potential imbalance of baseline confounders for each sub-trial. However, when PP analysis is of interest, IPC weights or consequent IPT weights are needed to address further selection bias caused by artificial censoring. In this case, (a) if weighting at baseline and subsequent IPC weights are used, weights used in outcome models are IPT weights (baseline) * IPC weights,[3] (b) if weighting at baseline and subsequent time-varying IPT weights are used, final weights are time-varying IPT weights that incorporate baseline, and (c) if matching at baseline is used, using only subsequent IPC weights (or IPT weights) is adequate. For the cloning approach, if all baseline confounders (even unmeasured) are balanced between the treatment and control groups at study entry, then, subsequent time-varying IPC weights or IPT weights are adopted to address artificial censoring. In this case there is no concern on combining methods dealing with confounders at baseline and artificial censoring thereafter.

In emulation analysis, multiple studies have adopted the sequential trial or cloning approaches to avoid potential IMT bias caused by mistakenly defined time zero.[24-27] Under the sequential trial strategy, each sub-trial is an emulated trial, and standard analysis methods can be applied. The enrollment period of each sub-trial (for example, a popular choice is one month) somewhat mimics grace period or landmark period during which treatment is initiated. Similarly, the cloning approach can be applied to account for grace period as for a clinical trial. When an ITT effect is of interest, the sequential trial approach adopts matching or weighting to balance baseline confounders, and no subsequent artificial censoring with IPC weighting or IPT weighting is needed. In contrast, ITT analysis with the cloning approach may be less straightforward. In particular, there are two possible versions. As introduced in Hernan et al. (2018)[28] and Petito et al. (2020),[26] one version is the original cloning idea with artificial censoring and subsequently updated weights during entire follow-up period, and the derived effect is claimed as the PP analog. The second version is to restrict that artificial censoring may only happen within grace period, and thus weights are no longer updated after grace period.[25] Although it is claimed that estimator is more like a PP effect, this approach does have an ITT flavor, as treatment switch that happens after grace period is ignored and adherence is assumed. Alternatively, the cloning idea realizes "randomization" for entire follow-up by artificial censoring and applies adherence or censoring weights until end of grace period (which mimics an ITT analysis) or end of follow-up (which mimics a PP analysis) to address unbalanced confounders and selection bias.

**A.5 Remarks**

When combined with causal inference, each method has its own suitable conditions, prerequisites, concerns, and limitations. Our recommendations are as follows: (a) If propensity score weighting or matching is performed, it is easier to adopt methods that only use time-independent propensity scores, such as methods with redefined time zero and the sequential trial approach; (b) With emulation, when an ITT effect is of interest, redefining time zero after grace period and the sequential trial approach can potentially get estimation with aligned interpretations using assigned treatment to impute missing planned treatment; and when a PP effect is of interest, the time-varying treatment modeling, sequential trial approach, and cloning approach with appropriate weighting are more appropriate.

**References**

1. D'agostino RB. Propensity score methods for bias reduction in the comparison of a treatment to a non-randomized control group. *Statist Med* 1998; 17: 2265-2281.
2. Brookhart MA, Wyss R, Layton JB, et al. Propensity Score Methods for Confounding Control in Nonexperimental Research. *Circulation: Cardiovascular Quality and Outcomes* 2013; 6: 604-611. DOI: 10.1161/CIRCOUTCOMES.113.000359.
3. Hernán MA, Brumback B and Robins JM. Marginal structural models to estimate the causal effect of zidovudine on the survival of HIV-positive men. *Epidemiology* 2000; 11: 561-570. 2000/08/24. DOI: 10.1097/00001648-200009000-00012.
4. Robins JM. Marginal Structural Models versus Structural nested Models as Tools for Causal inference. In: *Statistical Models in Epidemiology, the Environment, and Clinical Trials* (eds Halloran ME and Berry D), New York, NY, 2000, pp.95-133. Springer New York.
5. Robins JM, Hernán MA and Brumback B. Marginal structural models and causal inference in epidemiology. *Epidemiology* 2000; 11: 550-560. 2000/08/24. DOI: 10.1097/00001648-200009000-00011.
6. Hernan MA and Robins JM. Using Big Data to Emulate a Target Trial When a Randomized Trial Is Not Available. *Am J Epidemiol* 2016; 183: 758-764. 2016/03/20. DOI: 10.1093/aje/kwv254.
7. Liu R, Rizzo S, Whipple S, et al. Evaluating eligibility criteria of oncology trials using real-world data and AI. *Nature* 2021; 592: 629-633.
8. Admon AJ, Donnelly JP, Casey JD, et al. Emulating a novel clinical trial using existing observational data. predicting results of the PreVent study. *Annals of the American Thoracic Society* 2019; 16: 998-1007.
9. Caniglia EC, Robins JM, Cain LE, et al. Emulating a trial of joint dynamic strategies: An application to monitoring and treatment of HIV-positive individuals. *Stat Med* 2019; 38: 2428-2446. 2019/03/19. DOI: 10.1002/sim.8120.
10. Atkinson A, Zwahlen M, Barger D, et al. Withholding primary PcP prophylaxis in virologically suppressed HIV patients: An emulation of a pragmatic trial in COHERE. *Clin Infect Dis* 2020 2020/05/26. DOI: 10.1093/cid/ciaa615.
11. van Walraven C, Davis D, Forster AJ, et al. Time-dependent bias was common in survival analyses published in leading clinical journals. *Journal of Clinical Epidemiology* 2004; 57: 672-682. DOI: https://doi.org/10.1016/j.jclinepi.2003.12.008.
12. Stuart EA, Bradshaw CP and Leaf PJ. Assessing the generalizability of randomized trial results to target populations. *Prev Sci* 2015; 16: 475-485. 2014/10/14. DOI: 10.1007/s11121-014-0513-z.
13. St. Sauver JL, Grossardt BR, Leibson CL, et al. Generalizability of Epidemiological Findings and Public Health Decisions: An Illustration From the Rochester Epidemiology Project. *Mayo Clinic Proceedings* 2012; 87: 151-160. DOI: https://doi.org/10.1016/j.mayocp.2011.11.009.
14. Bernasconi DP, Rebora P, Iacobelli S, et al. Survival probabilities with time-dependent treatment indicator: quantities and non-parametric estimators. *Stat Med* 2016; 35: 1032-1048. 2015/10/28. DOI: 10.1002/sim.6765.
15. Hernán MA, Hernández-Díaz S and Robins JM. Randomized trials analyzed as observational studies. *Ann Intern Med* 2013; 159: 560-562. 2013/09/11. DOI: 10.7326/0003-4819-159-8-201310150-00709.
16. Hernán MA and Robins JM. *Causal Inference: What If*. Boca Raton: Chapman & Hall/CRC, 2020.
17. Stuart EA. Matching methods for causal inference: A review and a look forward. *Stat Sci* 2010; 25: 1-21. 2010/09/28. DOI: 10.1214/09-sts313.
18. Frölich M. Finite-Sample Properties of Propensity-Score Matching and Weighting Estimators. *The Review of Economics and Statistics* 2004; 86: 77-90. DOI: 10.1162/003465304323023697.
19. Robins JM and Finkelstein DM. Correcting for Noncompliance and Dependent Censoring in an AIDS Clinical Trial with Inverse Probability of Censoring Weighted (IPCW) Log-Rank Tests. *Biometrics*

2000; 56: 779-788. https://doi.org/10.1111/j.0006-341X.2000.00779.x. DOI: https://doi.org/10.1111/j.0006-341X.2000.00779.x.
20. Daniel RM, Cousens SN, De Stavola BL, et al. Methods for dealing with time-dependent confounding. *Stat Med* 2013; 32: 1584-1618. 2012/12/05. DOI: 10.1002/sim.5686.
21. Peduzzi P, Wittes J, Detre K, et al. Analysis as-randomized and the problem of non-adherence: an example from the Veterans Affairs Randomized Trial of Coronary Artery Bypass Surgery. *Stat Med* 1993; 12: 1185-1195. 1993/07/15. DOI: 10.1002/sim.4780121302.
22. Zheng C, Dai R, Gale RP, et al. Causal inference in randomized clinical trials. *Bone Marrow Transplantation* 2020; 55: 4-8. DOI: 10.1038/s41409-018-0424-x.
23. Murray EJ, Caniglia EC and Petito LC. Causal survival analysis: A guide to estimating intention-to-treat and per-protocol effects from randomized clinical trials with non-adherence. *Research Methods in Medicine & Health Sciences* 2021; 2: 39-49. DOI: 10.1177/2632084320961043.
24. Gupta S, Wang W, Hayek SS, et al. Association Between Early Treatment With Tocilizumab and Mortality Among Critically Ill Patients With COVID-19. *JAMA Intern Med* 2021; 181: 41-51. 2020/10/21. DOI: 10.1001/jamainternmed.2020.6252.
25. Maringe C, Benitez Majano S, Exarchakou A, et al. Reflection on modern methods: trial emulation in the presence of immortal-time bias. Assessing the benefit of major surgery for elderly lung cancer patients using observational data. *Int J Epidemiol* 2020; 49: 1719-1729. 2020/05/10. DOI: 10.1093/ije/dyaa057.
26. Petito LC, García-Albéniz X, Logan RW, et al. Estimates of Overall Survival in Patients With Cancer Receiving Different Treatment Regimens: Emulating Hypothetical Target Trials in the Surveillance, Epidemiology, and End Results (SEER)-Medicare Linked Database. *JAMA Netw Open* 2020; 3: e200452. 2020/03/07. DOI: 10.1001/jamanetworkopen.2020.0452.
27. Danaei G, Garcia Rodriguez LA, Cantero OF, et al. Electronic medical records can be used to emulate target trials of sustained treatment strategies. *J Clin Epidemiol* 2018; 96: 12-22. 2017/12/06. DOI: 10.1016/j.jclinepi.2017.11.021.
28. Hernán MA. How to estimate the effect of treatment duration on survival outcomes using observational data. *BMJ* 2018; 360: k182. DOI: 10.1136/bmj.k182.